\documentclass[12pt,preprint]{aastex}
\slugcomment{To appear in The Astrophysical Journal}

\def\tighttable{\def\baselinestretch{1.0}}
\def\kms{${\rm km\,s}^{-1}$}

\def\ergscm{${\rm ergs\,s}^{-1}\,{\rm cm}^{-2}$}

\newcommand {\lya}{Ly$\alpha$}
\newcommand {\lyb}{Ly$\beta$}
\newcommand {\ciii}{\ion{C}{3}}
\newcommand {\civ}{\ion{C}{4}}
\newcommand {\niii}{\ion{N}{3}}
\newcommand {\nii}{\ion{N}{2}}
\newcommand {\ovi}{\ion{O}{6}}
\newcommand {\oiii}{\ion{O}{3}}
\newcommand{\he}{\ion{He}{2}}
\newcommand {\lm}{$\lambda$}

\newcommand {\ngc}{NGC 1068}
\shorttitle{NGC 1068}
\shortauthors{Zheng et al.}

\begin{document}

\title{Spatially Resolved Far-Ultraviolet Spectroscopy of\\the Nuclear Region 
 of NGC 1068
\altaffilmark{1}}

\author{
Wei Zheng\altaffilmark{2}, 
Jun-Xian Wang\altaffilmark{2,3}, 
Gerard A. Kriss\altaffilmark{2,4}, 
David Sahnow\altaffilmark{2},
Mark Allen\altaffilmark{5},
Michael Dopita\altaffilmark{6},
Zlatan Tsvetanov\altaffilmark{2,7}, 
and
Geoffrey Bicknell\altaffilmark{6}
}
\altaffiltext{1}{
Based on observations made with the NASA-CNES-CSA Far Ultraviolet 
Spectroscopic Explorer (FUSE), which is operated for NASA by the Johns Hopkins
University under NASA contract NAS5-32985,
and observations with the NASA/ESA {\it Hubble Space Telescope}, obtained at 
the Space Telescope Science Institute, which is operated by the Association of 
Universities of Research in Astronomy, Inc., under NASA contract NAS5-26555.}
\altaffiltext{2}{Center for Astrophysical Sciences, Department of Physics and
 Astronomy, The Johns Hopkins University, Baltimore, MD 21218}
\altaffiltext{3}{Center for Astrophysics, University of Science and Technology 
  of China, Hefei, Anhui, 230026, China} 
\altaffiltext{4}{Space Telescope Science Institute, 3700
 San Martin Drive, Baltimore, MD 21218} 
\altaffiltext{5}{
Observatoire de Strasbourg, 
67000  Strasbourg, France
}
\altaffiltext{6}{
Mount Stromlo \& Siding Spring Observatories, Australian National University, 
ACT 2611, Australia
}
\altaffiltext{7}{NASA Headquarters, Washington, DC 20546-0001}

\begin{abstract}
We carry out high-resolution FUSE spectroscopy of the nuclear region of NGC 1068.
The first set of spectra was obtained with a 30\arcsec\ square aperture that
collects all emission from the narrow-line region. The data reveal a strong 
broad \ion{O}{6} component of FWHM $\sim 3500$ $\rm km s^{-1}$ and two narrow 
\ion{O}{6} $\lambda\lambda 1031/1037$ components of $\sim 350\ \rm 
km s^{-1}$. The \ion{C}{3} $\lambda 977$ and \ion{N}{3} $\lambda 991$ 
emission lines in this spectrum can be fitted with a narrow component of 
FWHM $\sim 1000$ $\rm km s^{-1}$ and a broad one of $\sim 2500$ $\rm km s^{-1}$. 
Another set of seven spatially 
resolved spectra were made using a long slit of $1\farcs 25 \times 20 \arcsec$,
at steps of $\sim 1$\arcsec\ along the axis of the emission-line cone.
We find that (1) Major emission lines in the FUSE wavelength range consist of 
a broad and a narrow component; (2) There is a gradient in the velocity field 
for the narrow \ion{O}{6} component of $\sim 200$ $\rm km s^{-1}$ from 
$\sim 2 \arcsec$ southwest of the nucleus to $\sim 4 \arcsec$ northeast. 
A similar pattern is also observed with the broad \ovi\ component, with a 
gradient of $\sim 3000\ \rm km s^{-1}$. These are consistent with the HST/STIS
findings and suggest a biconical structure in which the velocity field is mainly radial
outflow; 
(3) A major portion of the \ion{C}{3} and \ion{N}{3} line flux is produced in 
the compact core. They are therefore not effective 
temperature diagnostics for the conical region; and (4) The best-fitted UV 
continuum suggests virtually no reddening, and the \ion{He}{2} 1085/1640 ratio 
suggests a consistently low extinction 
factor across the cone. At $\sim 2 \arcsec$ northeast of 
the nucleus there is a region characterized by (a) a strong Ly$\alpha$ flux, but normal 
\ion{C}{4} flux; (b) a broad \ion{O}{6} line; and (c) a significantly enhanced \ion{C}{3} 
flux. 
\end{abstract}


\section{INTRODUCTION}

NGC 1068 is a prototypical Seyfert 2 galaxy. Because of its proximity 
($z=0.0038$) and brightness, it has been studied in nearly every possible 
detail. The polarimetric observation by \citet{antonucci}, which reveals
a Seyfert 1 spectrum in scattered light, suggests that the
nucleus and its associated broad-line region (BLR) are obscured. 
This finding provides strong evidence for the unified theory in which viewing 
angles account for the differences between various active galactic nuclei
\citep{antonucci2}

The nuclear region of NGC 1068 harbors a variety of astrophysical phenomena. 
At the very center of the nucleus there is a bright compact ($<$0\farcs 3) region
commonly referred as ``the hot spot''. 
Within a few arcseconds from the nucleus, there are several bright and compact 
clouds that coincide with knots in the radio jets \citep{wilson,evans}. 
The narrow-line region (NLR) is conical in shape toward the northeast (Fig. 1), 
along a position angle of $\sim 200 \deg$ and with an opening angle 
of $\sim 40 \deg$.
Beyond a 6\arcsec\ radius the surface brightness drops dramatically, and emission 
is dominated by two ring-like filaments at $\sim10$\arcsec\ and 
15\arcsec\ from the nucleus. 

High-spatial-resolution spectroscopy of NGC 1068 has been carried out in the
optical \citep{caganoff,unger,inglis,emsellem,gmos} as well as in the UV 
\citep{caganoff,krm2,krm,cecil,groves}. From approximately 2\arcsec\ southwest of the nucleus 
to 4\arcsec\ northeast, 
emission lines exhibit multiple components \citep{cecil90,krm0}: (1) major 
emission lines consist of narrow and broad lines; (2) broad lines are approximately 
2500-4000 \kms\ wide, which may be linked to those that are found in polarized light and 
believed to be reflected light from the inner BLR, and (3) narrow lines 
consist of a pair of red and blue components. The [\oiii] and [\nii] line profiles suggest 
that the separation of these two components varies across the conical NLR. 
In addition to an overall 
biconical ionization configuration, there are compact knots whose optical spectra  
resemble kinematically the associated absorption line systems in quasars 
\citep{cecil,ckg,krm4}. 
These line-emitting knots 
have blueshifted radial velocities up to 3000 \kms\ relative to the galaxy's 
systemic velocity, contributing mostly to the emission-line flux but not the continuum.
Between $\sim$2\farcs 5 and 4\farcs 5 northeast from the nucleus, UV line emission is 
redshifted relative to the systemic value, a pattern that is interpreted as the
 expansion of the plasma in the radio lobe \citep{axon}.

Several important emission lines in the far-UV (FUV) region between 912 
and 1150~\AA\ are observable only with specially crafted UV instruments.
During the Astro-1 mission the Hopkins Ultraviolet Telescope (HUT)
observed NGC 1068 with 18\arcsec\ and 30\arcsec\ apertures.
The most striking features in the wavelengths below 1150 \AA\ 
are the strong \ciii\ \lm 977 and \niii\ \lm 991 lines.
The line intensity ratios of \ciii\ I(\lm 1909)/I(\lm 977) and 
\niii\ I(\lm 1750)/I(\lm 991) are temperature sensitive, and the derived 
temperature is $> 25$ 000 K \citep{kriss}, higher than the values expected for a region 
producing \ion{C}{3} and \ion{N}{3} emission by photoionization,
The line ratios in \ngc\ 
are similar to those of the Cygnus Loop supernova remnant \citep{blair},
suggesting a significant contribution from shock-heating mechanisms. 
Astro-2 observations of NGC 1068 were obtained with a 12\arcsec\ aperture 
at three different positions. The results \citep{grimes} suggest
that the emission lines observed with HUT likely arise in the inner nuclear region 
imaged with HST. However, the poor angular resolution of the HUT instrument
($> 10\arcsec$) does not allow a study of the NLR in terms of spatial details.

The line-emitting mechanisms in the NLR, {\em i.e.} photoionization from
the nucleus or shocks produced by jets, have long been under debate.
\citet{dopita} and \citet{bicknell} proposed that 
emission in the NLR may be entirely caused by shocks. 
Velocity splitting over 1000 \kms, reported by \citet{axon}, 
in the vicinity of some of the bright emission-line 
knots provides evidence that fast shocks exist in 
the NLR of NGC~1068. However, Seyfert galaxies host powerful nuclear 
sources of ionizing radiation, and the situation
can be more complex with shocks or nuclear photoionization prevailing
in different environments \citep{allen,morganti}.
\citet{morse} suggested that 
photoionizing shocks are important when a radio jet interacts with the 
interstellar medium, but not in the objects where sharp, straight-edged 
ionization cones are observed. \citet{ferguson} 
argued that strong \ciii\ \lm 977 and \niii\ \lm 991 emission may arise from 
fluorescence in photoionized gas if turbulent velocities exceed 
$\sim$1000 \kms.  However, \citet{grimes} found that such a 
velocity would lead to extreme physical conditions that are inconsistent with 
the Astro-2 data. More recent HST data \citep{krm,cecil} found that the emission-line 
ratios  are consistent with photoionization instead of shock heating mechanism. 

We have carried out observations with the {\em Far Ultraviolet
Spectroscopic Explorer} (FUSE) to study the spatial distribution 
of FUV emission lines. In this paper we present the results of both large-aperture
and spatially resolved spectroscopic observations with FUSE. For the first time, we 
are able to study the position dependence of several important diagnostic lines 
in the FUV band.

\section{DATA}

\subsection{FUSE data}

FUSE covers a wavelength range between 904 and 1188 \AA, with a 
spectral resolution of $R \sim 20 000$ \citep{moos}. It is
based on a Rowland circle design and consists of four separate optical paths 
or channels. Each channel consists of a mirror coated with aluminum plus LiF 
or SiC, a focal plane assembly, which includes the spectrograph apertures, a 
diffraction grating coated with aluminum plus LiF or SiC, and a portion of 
FUV detector (two named A and B). 
In all, data are collected from eight traces, and the pairs of A and B are 
called channels. Our original plan was to
take spectra of the nuclear region of \ngc\ at seven positions separated by 
$\sim$ 1\arcsec.
Because of thermal instability of the optical structure in orbit, the four 
channels in the FUSE spectrograph are not perfectly aligned, with 
orbit-dependent drifts possibly as large as 6\arcsec. 
Therefore, a new observing strategy was 
developed to counter the drifting effect.
The high-resolution aperture 
(HIRS: $1\farcs 25 \times 20 \arcsec$) is used for spectroscopy on the finest
spatial scales, to ensure that maximum 
resolution is maintainable even if the telescope imaging or pointing 
stability degrades below specifications. 
Since the LiF1 channel is mounted on the same optical system as the FUSE 
fine error sensor, the spatial knowledge and stability is fully available 
for the data from this channel.

The observations of \ngc\ were carried out with two modes between 2001 
November 28 and December 02. The first set of data of 21951 seconds
was taken with fixed pointing and the LWRS (low-resolution) square aperture 
of 30\arcsec. With such a 
large aperture, fluxes of the nuclear region are all collected (Fig. 2), 
even with the thermally-induced mirror motions anticipated. 
The second set of observations was carried out in a scanning mode across 
the conical emission-line region, with a total exposure of 94434 seconds.
A narrow HIRS slit at a position angle of $137 \deg$ was used. 
Because of orbital constraints, we were not able to place the slit 
completely perpendicular to the conical axis. The slit actually makes 
an angle of $\sim 97 \deg$ with respect to the conical axis.

Over years of FUSE operation, a large amount of engineering data have been 
collected, 
and the drift pattern among the four channels has been well studied. After
peak-up alignments of channels with a bright point source at the beginning 
of orbital nights, the relative drifts exhibit a dependence on orbital time 
(Fig. 3). The LiF2 channel exhibits relatively small drifts, within one 
arcsecond
with respect to the LiF1 channel. The SiC channels exhibit drifts that can be 
represented by a constant drift rate during 
orbital nights. For example, the SiC2 channel drifts relative to the LiF1 
channel at a rate of approximately 4\arcsec\ per 30 minutes.
The starting drift at the orbital sunset is however variable, depending on 
the degree of occultation because the drift is generally in the reverse 
direction during orbital days.

NGC 1068 itself is too diffuse to appear in the guide-star camera. 
For target acquisition, we used an offset guide star $\sim 85$\arcsec\ from 
the nucleus. We also used a FUV-bright star, Feige 23, which is approximately 
four degrees away from the source, for instrumental peak-up alignments at 
the beginning of each pointing.
One orbit of peak-up operation on Feige 23 was made to establish the drift 
pattern due to orbital motion. During the second orbit, one peak-up with the 
MDRS aperture ($4\arcsec \times 20$\arcsec) and another with HIRS were carried 
out on Feige 23. 
The instrument was then pointed to the southwest part of the nucleus of 
\ngc, and we started observations in a scanning mode. 
As shown in Table 1, each of the five pointings consists of 7-9 orbits on the 
target. The observation window varied orbit by orbit, between $\sim 1200$ and
2600 sec. The scanning speed in each individual orbit was set to complete 
10\arcsec\ within the designated observation window.
After accumulating a total exposure time of $\sim 18000$ sec, FUSE 
returned to Feige 23 and carried out one peak-up with MDRS and another with 
HIRS, then moved back to \ngc\ for a new set of scans. After a total of 
five pointings, FUSE returned to Feige 23 for one last orbit of peak-up with MDRS, to 
acquire additional information on the drift pattern.

Since the channel alignments are not perfect except at the beginning of the 
first orbit, the orbit-time dependence of channel drifts derived from the 
engineering data is only useful in terms of the relative drift speed, not for
the absolute timing.

The drifts in the SiC channels can be as large as 6\arcsec, which is 
near the size of the ionization cone. The drift direction  
is mainly along +x on the FUSE aperture plate, i.e., along the conical 
direction of NGC 1068 at the desired FUSE aperture position angle of $\sim
105 \deg$. 
Therefore the observation procedure was designed so that each observation 
starts $\sim 5\arcsec$ from the nucleus in the anti-cone (southwest) direction 
and sweeps across the cone towards the northeast direction at a constant 
angular velocity. 
Fig. 4 illustrates the time sequence in one orbit of observation: with a constant 
angular speed, 
the FUSE HIRS slit swept across the nuclear region of NGC 1068. We identified the peak
position of the count rate in the LiF1 channel as the nuclear position.
Based on the angular speed illustrated in Fig. 3, seven time intervals were defined,
and the spectral extraction in these segments yields the data A-G.
The upper panel of Fig. 4 marks the segments for the LiF1 channel, whose spatial information
is fully available. For the other three channels, their count rates peaked
at different times as a result of drifting. We marked their 
respective peak positions. The sweeping speeds in these channels are the sum of
two terms: that of the LiF1 channel plus the relative drift speed of each 
channel (Fig. 3) with respect to LiF1. Their time segments A-G are also 
illustrated in Fig. 4. 
The observation results in a series of segments that are dependent on the 
relative positions with respect to the nucleus. In each time sequence, one can 
determine the time when the observation slits coincide with the nucleus, as
determined by the peak in the count rate. 

The LiF1 slit positions are illustrated in Fig. 1. Seven sets of data are 
extracted with a relative shift of 1 arcsec (1\farcs 25 at the nucleus), 
starting from 2\arcsec\ southwest from the nucleus
(in the opposite direction of the cone), to $\sim 4\arcsec$ northeast of it. 
They were named as A,B,C (the nucleus, see Fig. 1), D,E,F and G.
The significant astigmatism of FUSE yields a spatial resolution of 
$\sim 5\arcsec$ at best, therefore it cannot be used like a normal long-slit 
spectrograph to dissect the fine details along its slit. Not all the segment data 
are extracted: If a time segment is incomplete (``G'' in Fig. 4, for example),
they are not used.

Four pairs of spectra were derived from the FUSE channels. Each pair of  ``A'' 
and ``B'' data is subject to the same normalization factor. We used the \ovi\ emission 
flux, which is common to all four channels, 
to normalize the spectra to the level in LiF1A. We first calculated the continuum level 
between 1050 and 1070 \AA\ for LiF2B, SiC1A, SiC 2B and LiF1A, then subtracted it from the 
spectrum to enable the measurement of emission-line fluxes. We then calculated the 
\ovi\ flux in two bins: a narrow component between 1041.1 and 1043.6 
\AA, and a broad component between 1027 and 1048 \AA. The flux ratios with respect to their
values in the LiF1A spectrum were calculated, then their mean values were used 
as the normalization factors (Table 2).
For the first set of data, taken with LWRS, we corrected the 
``worm effect'' in the LiF1A,B channel, according to the procedures described 
in the {\it FUSE Instrument and Data Handbook} 
(http://fuse.pha.jhu.edu/analysis/dhbook.html).

FUSE data reduction was carried out with pipeline version CALFUSE2.3, with specific timing 
flags to extract time-dependent (hence position-dependent) segments of 
spectra in eight traces. 
The FUSE spatial 
resolution is 1\farcs 5, and the spacecraft jitter, as determined from the FUSE 
engineering data, is approximately 0\farcs 6. The total spatial resolution is 
therefore 1\farcs 6. Since the data were obtained with a 1\farcs 25 slit,
there is overlap between positions. This can be seen by the fact that the sum 
of the fluxes in all positions is larger than that taken with one large 
30\arcsec\ aperture. 

The normalized spectra in the four channels are binned to 0.1\AA, then merged, 
weighted by their signal-to-noise ratio. The merged spectra at the 
seven slit positions are plotted in Fig. 5. The wavelength bin between $\sim 1077$ 
and 1087 \AA\ falls into a gap between the LiF channels, and only data in the 
SiC2B channel exist.

\subsection{HST data}

We retrieved archival HST/STIS spectroscopic data to complement our FUSE data.
Two sets of UV spectra are available: one taken at a position angle of $218 
\deg$ and four adjacent slit positions of $0\farcs 2 \times 52$\arcsec\ 
\citep[][STIS-A hereafter]{cecil}, and another at a position angle of 
$202 \deg$ taken with a slit of $0\farcs 1 \times 52$\arcsec\ \citep[][STIS-B]{krm}. 
Both sets (see Fig. 1) contain data taken with grating G140L and G230L, 
covering a wavelength range of $\sim 1150$ to 3170 \AA\ at a resolution of $R \sim 1000$.
These STIS observations were made between 1988 and 2000, therefore 
not simultaneous with the FUSE data.
For STIS-A, we only used three slit positions as the other does not have G140L 
data.  The fluxes in the three extracted spectra are summed up.

Since the FUSE spatial resolution ($\ge 5\arcsec$ along its slit) is considerably 
lower than that of STIS, and 
since the position angles of these two sets of observations are nearly 
perpendicular, we must normalize the STIS data to that of 
FUSE. Using the continuum level around 1180 \AA\ in these spectra 
did not result in
satisfactory matches as the line/continuum ratios are not constant at these spatial scales 
across the nucleus. We therefore use HST WPFC2 images retrieved from MAST to determine the 
normalization factor between STIS and FUSE data. 

To extract the STIS spectra, we first smoothed the 2-D images by 30
pixels (0\farcs 67) with a Gaussian kernel along the spatial direction of 
the FUSE scans, to 
match the spatial resolution of the corresponding FUSE data. We then ran the 
standard STIS pipeline task ``x1d''
with steps of one arcsecond (1\farcs 25 at the nucleus) that correspond
to the respective moving FUSE slit positions, and a slit width of 1\farcs 25.
To normalize the STIS fluxes, we used a HST/WFPC2 image of \ngc, taken with 
filter F218W, and smoothed it along the same direction with the same kernel size 
as the STIS 2-D image. We then measured the flux in each STIS window. 
To normalize the FUSE fluxes, we
smoothed the WFPC2 image to the FUSE resolution, namely 0\farcs 67 and 5\farcs,
respectively, along the direction perpendicular and parallel to that of the 
FUSE slit (Fig. 1). The flux ratios in respective extraction windows are used 
to derive the normalization factors (Table 2). To the northwest direction from
the source, there are several  bright spots whose fluxes may be picked up 
because of FUSE's low spatial resolution along its slit. We tested with several
extraction windows that are sufficiently short not to include the fluxes from 
these bright spots in the smoothed WFPC2 image. Since it is 
unlikely that these bright spots contribute to the redshifted UV emission lines, we used 
the flux values measured in a short window in calculating the scaling factor. 
Any uncertainties introduced by this window selection only affect the scaling 
of \lya\ and \civ\ in position A and G.
 
Since the FUSE slit positions are accurate to $\sim$ 0\farcs 3, we measured 
the fluxes in the neighboring regions, to make sure that the
normalization factors are accurate to at least 20\%.
The height of STIS extraction windows is the same as that of FUSE 
($\sim 1$\arcsec), but their widths (0\farcs 1-0\farcs 2) are much smaller than
FUSE (20\arcsec). Therefore considerable STIS normalization factors were 
applied. The matching STIS spectra are plotted in Fig. 6.

We compared extracted spectra from the STIS-A and STIS-B datasets, and their 
line ratios are comparable within 25\%. In the following section, we only use 
the data from STIS-A, as they cover a spatial region six times as wide ($0\farcs 2 
\times 3 \arcsec$) as that of STIS-B.

\section{FITTING}

Spectral analyses were carried out using IRAF task {\it Specfit}
\citep{specfit}.
For the FUSE spectrum taken with a large aperture, we used a pair of narrow components 
and one broad component in the \ovi\ emission line profile. Each narrow \ovi\ component 
was modeled as a doublet whose wavelength ratio is fixed by atomic data and whose  
line widths are identical.
The \ovi\ emission feature is heavily absorbed by interstellar absorption 
at $\sim$ 1037\AA, therefore we introduced several absorption components. The \lyb\
emission is not prominent, and it is often overwhelmed by the broad \ovi\ emission. 
Only in one or two positions is the \lyb\ emission visible. We therefore only 
modeled one component with a fixed line width. For \niii\ and \ciii\ 
emission line features, one narrow 
component and one broad component were included in fitting. We fitted a power law with no 
extinction to the continuum. The fitted results are listed in Table 3.

For the FUSE spectra taken at positions A-G, we used one narrow component plus 
one 
broad component to model the \ovi\ emission, and one component for the \niii\ and \ciii\ 
lines. To reduce the effect of channel drifts, the FUSE spectra were fitted
individually for each channel. The fitting results in Table 4 are primarily 
derived from the LiF1 data, which are not affected by channel drifts. For 
wavelengths below 995 \AA, we used the fitting results from the scaled SiC2A 
channel, which extends to 
1005 \AA\ and is therefore less susceptible than the SiC1B channel to the detector
edge effects. For \ion{He}{2} 1085, LiF2A data were used.
The fluxes of fitted components were normalized by the values in Table 2, and 
then tabulated in Table 4. The profiles of four major emission lines at 
seven different slit positions are plotted in Fig. 7, to show the changes in 
the narrow and broad components as a function of spatial location.

The UV spectra suggest a low extinction level. We first fitted the continuum 
in the wavelength windows that are free of emission 
and absorption lines, and the results suggest a power-law index of 
$\alpha = 0.90 \pm 0.04$ ($f_\lambda \propto  \lambda^\alpha$) 
and an extinction of $E_{B-V}=0.005 \pm 0.002$. As a comparison, the HUT 
spectrum between 912 and 1800 \AA\ was fitted with a power-law continuum of 
$\alpha=0.62 \pm 0.15$ and $E_{B-V}=0.065 \pm 0.02$. If we adopt this 
extinction value, the power-law index is $\alpha = 0.85 \pm 0.06$. The 
similar power indices suggest that one single power-law continuum can fit the 
far-UV spectrum of NGC 1068 without introducing significant reddening,  
as suggested from the UV/optical line ratios \citep{krm0}.

\section{Discussion}

Many previous studies have gradually unraveled the kinematical 
complexities of gas in the nuclear region of NGC 1068. The X-ray data 
\citep{xmm} suggest that emission lines are formed mainly in a photoionied 
plasma of a temperature around a few eV. The spatial resolution afforded by 
HST leads to a picture of decelerating jet \citep{das}: 
a biconical outflow from the nuclear region sweeps up denser, ambient 
clouds in the interstellar medium (ISM) of NGC 1068. Other possibilities
include overlapping, discrete ejection that gradually dissipates 
\citep{axon,capetti}. In addition, a high-velocity radio jet 
impinges on some of the clouds. Some gas expands perpendicularly to the 
axis of the jet, and the expanding radio lobe at the end of the jet also 
pushes on the ambient ISM. Near the nucleus, kinematic components span 
several thousand kilometers per second in velocity, and the continuum 
hot spot visible in HST images reflects a polarized view of the broad 
lines in the active nucleus. 

Our spatially resolved FUSE observations, while not at the resolution of 
HST, add information from major emission lines shortward of the HST 
bandpass at high spectral resolution. We plot the fluxes of major UV 
emission lines from our observations at seven slit positions in Fig. 8, 
and list the fitted line properties in Table 4. In this section, we discuss 
the four emission lines in the FUSE spectral range, along with a comparison to 
emission lines observed in the STIS spectra.

\subsection{\ovi\ emission} 

The most prominent feature in the FUSE spectra is the \ovi\ \lm\lm 1031,1037 
emission line. In the data taken with LWRS, the two narrow \ovi\ 
components are of FWHM $\sim 350$ \kms\ and a separation of $\sim 200$ \kms. 
In the 
seven spectra taken with a narrow slit, we resolve this blend into one
narrow component with FWHM of $\sim 350$ \kms\ and one broad component.
In Fig. 9 we plot the \ovi\ profiles at the seven different slit positions.
The high spectral resolution of FUSE data enables us to compare with the 
results of optical Fabry-Perot spectroscopy \citep{cecil90}, which reveal 
a narrow core of $\sim 300$ \kms. 
According to the optical data, nearly 75\% of the [\nii] \lm 6583 flux is from 
components of $\sim 1500$ \kms\ wide. Line widths at such a scale are 
consistent with that derived from HST UV spectroscopy. 
The narrow-line flux in the FUSE spectra is highly concentrated (60\%) 
at the compact core, suggesting that 
the ``true'' NLR probably remains unresolved, at a sub-arcsecond scale.
While the narrow O VI emission line is not the 
dominant component, its distribution is different from its broad 
counterpart, while only 40\% of the broad \ovi\ line flux is from 
slit position C.

It is surprising that the \ovi\ emission is dominated by a component 
that is broader than those seen in polarized light: in the LWRS spectrum, more 
than 3/4 of the \ovi\
flux is from a component with FWHM $\sim 3500$ \kms\ that is blue-shifted 
relative to the narrow component by $\sim 500$ \kms. This broad component 
is present in
all seven FUSE slit spectra with considerable strength. It may arise from the 
reflected emission from the hidden BLR, and/or may be the result of a 
significant velocity dispersion in the NLR. The HST FOC data \citep{axon} 
reveal that emission
lines near the hot knots 2\arcsec\ northeast (FUSE slit positions D and E) 
are split into two velocity systems separated by $\sim 1500$ \kms.
The STIS spectra discussed by \citet{groves} show [\oiii] emission 
knots spanning such a broad velocity range in the immediate vicinity of 
the nucleus, but not at distances of several arcseconds.
Since the FUSE slit collects emission from a block of regions spanning 
several arcseconds perpendicular to the conical axis, the total line 
emission from these regions may be blended into one broad component.

The ratios of \civ/\ovi\ may provide insight into the physical conditions of the 
line-emitting regions. The value is higher for the narrow components than their
broad counterparts, implying a range of the ionization parameter $U \sim 
0.05-1.0$ in a typical photionization calculation. High values of $U>1$ 
are consistent with models that assume the same origin for the 
associated absorbers and BLR \citep{n7469}, suggesting that the clouds that produce 
the broad emission components may be of the same origin as associated absorbers.

\subsection{Velocity field}

The narrow \ovi\ line exhibits a systematic velocity shift from position 
A to G by approximately
220 \kms. This gradient in the spatially resolved spectra explains why 
there are two narrow-line components in the integrated flux from the 
large-aperture FUSE spectrum, where it is unresolved.
\citet{krm0} reported a similar velocity pattern in their 
HST spectra. \citet{das} successfully modeled this as a biconical 
outflow in which radial velocity changes as a function of the distance 
to the central nucleus: the emission line knots show evidence for radial 
acceleration
to a projected distance of 2\arcsec\ to the northeast direction, followed by 
deceleration up to 4\arcsec. The \ovi\ line widths also increase at 
$\pm 2$\arcsec\ from the nucleus, probably implying a larger dispersion in 
these regions.

The broad component of \ovi\ exhibits a qualitatively similar kinematic 
pattern, but at a larger amplitude: its centroid shifts by $\sim 1500$ \kms\ 
(Fig. 10) across the same spatial region.
Fig. 11 shows similar trends for the narrow and broad components of \civ\ 
in the HST spectra which have not been explicitly noted in previous 
studies: the narrow component follows the kinematic pattern of the 
optical lines modeled by \citet{das}; the broad component of \civ\
shows behavior similar to that of \ovi\ in the FUSE spectra.
Prior observations that noted this blue-shifted broad component 
invariably attributed it to the reflection of the BLR. The 
blue shift and width (at the position of the hot spot, and in 
integrated light) are comparable to the broad polarized H$\beta$ line 
observed by \citet{antonucci}. 

In a scattered BLR picture, a blue shift is caused by the 
outflowing wind from the torus along our line of sight, and a line width is 
due to the intrinsic broad line width convolved with the thermal width 
of the hot reflecting wind. As a broad component is present in the FUSE spectra at all
seven positions, it is possible to assume that this is reflected light from the hidden BLR. 
However, a large covering factor is needed to 
explain the observed fluxes. Assuming a covering factor of 0.1, the intrinsic
flux of the broad \ovi\ emission in NGC 1068 would exceed that in NGC 4151. 
A more reasonable explanation for the observed broad line widths is the large 
velocity dispersion between bright knots.
At approximately 2\arcsec\ from 
the core (position A and E), the FWHMs of broad components are the 
broadest at 3200 \kms. These maxima coincide with the widest spliting of 
velocity in bright knots \cite{krm0}. With an intrinsic dispersion of 
$\sim$800 \kms\ and a separation in velocity of $\sim$2500 \kms\ between bright 
knots (from STIS results), data collected by the long FUSE slit would 
exhibit a broad component of $\sim$3200 \kms, which is what observed at 
positions A and E. Extended regions of hot, photoionized 
gas are seen in X-ray images of NGC 1068 \citep{young,xmm} that could be 
visible manifestations of this hot outflow. 
It is natural to assume that these high-velocity clouds may be related to the 
associated absorbers \citep[][and references therein]{ckg} in AGN, which are 
mostly blueshifted. 
As with the lower-velocity, lower ionization emission-line gas, the 
acceleration of this high ionization gas eventually is brought to a halt 
by an unknown deceleration mechanism, which might plausibly be 
interaction with the ambient ISM of NGC 1068. The evidence for deceleration 
at arcsecond scale may suggest that acceleration of the outflow materials may 
take place at sub-arcsecond scales.

\subsection{\ciii\ \lm 977}

The flux ratio of \ciii\ \lm 1909 to 977 is extremely sensitive to temperature, and the 
value measured in N1068 suggests a high temperature that is consistent with shock-heating 
\citep{kriss}. The FUSE data taken with a large aperture (Table 3) reveal that this emission
line consists of a narrow and a broad component. The broad component is weak and 
hence cannot be well separated in the data segments taken with a narrow FUSE slit.
The STIS data reveal that \ciii\ \lm 1909 emission can be fitted with a narrow and a broad 
component of FWHM $\sim 900$ and 3500 \kms, respectively. In principal, the ratio of \ciii\ 
\lm 1909 to 977 should be calculated only between the narrow components.

As shown in Fig. 8, the flux of \ciii\ \lm 977 is highly concentrated in position C and D.
High temperatures implied by this line emission may therefore be associated with the compact core. In a large portion of the ionization cone, the line ratios (\lm 1909/\lm 977) are 
considerably higher, suggesting a lower temperature. However, the line ratios 
are considerably lower than that derived from the HUT data, where broad \ciii\ 
\lm 1909 component was included in the calculation.

\subsection{\niii\ \lm 991}

The line ratio of \niii\ \lm 1750 to \lm 991 is also temperature dependent, and its value has 
also been used to derive a high temperature in the ionization cone. The FUSE data also
reveal a pair of components in the data taken with a large aperture. Unlike \ciii\ \lm 977,
the broad component of \niii\ \lm 991 , as shown in Table 3, is stronger than its narrow 
counterpart. The FUSE data taken with a narrow slit shows that nearly a half of this broad 
component is in position C. In other positions, the \niii\ \lm 991 emission can only be 
fitted with one component of FWHM $\sim 1000$ \kms. In combination  with the trend for 
\ciii\ \lm 977, we conclude that a bulk of flux in these temperature-sensitive emission lines 
is from the compact core of NGC 1068, and their intensity is not directly tied to the physical 
conditions in the ionization cone.

\subsection{\he\ 1085}

The distribution of \he\ \lm 1085 is different from other emission lines in the FUSE spectra: 
as shown in Fig. 8, the \he\ flux varies smoothly across the NLR region, like that of \lya, 
\civ\ and other lines in the STIS spectra. Since \he\ emission is believed to be produced mainly by 
recombination and is insensitive to the gas temperature, this pattern of variation may simply 
reflect the distribution of NLR gas across the ionization cone. 
The ratio of \he\ emission \lm 1640 to \lm 1085 is believed to be a reddening 
indicator. As shown in Fig. 12, the ratio is nearly constant ($\sim 5.5$) 
between slit 
positions B and F. This is consistent with the HUT result 
of $5.8 \pm 1.6$, suggesting an insignificant level of extinction 
($E_{B-V} \le 0.05$). The low values at positions A and G may suggest that 
the data at these positions are not reliable.

The \he\ I(1640)/I(1085) ratio is slightly lower than that expected 
from recombination, and this may suggest that the \he\ \lm 1085 may be 
slightly contaminated by other weak UV emission lines. 
\ion{N}{2} \lm 1085 is a likely candidate, and the trend of decreasing 
\he\ (1640)/I(1085) ratio with distance from the nucleus supports this as 
there is a noticeable decrease in ionization state at larger distances 
from the nucleus \citep{axon,krm2,krm3}.
As with \ovi\ and \civ, there appears to be a broad component
that is considerably blueshifted. However, the relevant wavelength range 
is in a gap between the LiF channels, and only SiC2B data are 
available. Because of the low 
S/N ratio, the reality of such a broad component is still questionable, 
as no such counterpart is found in the corresponding HST spectra of \he\
\lm 1640.

\section{Summary}

We have carried out high-spectral resolution, spatially resolved spectroscopy of 
the nuclear region of the Seyfert 2 galaxy NGC 1068. The high spectral resolution 
of FUSE data enables us to study the line profiles of \ion{O}{6}, \ion{N}{3}, 
and \ion{C}{3} as a function of position across the nuclear region. Our observations 
using a long slit of $1\farcs 25 \times 20 \arcsec$ at steps of $\sim 1 \arcsec$ 
along the axis of the emission-line cone resulted in a set of seven spatially 
resolved spectra running from  $\sim 2 \arcsec$ southwest of the nucleus to 
$\sim 4 \arcsec$  northeast.

The \ion{O}{6} profiles exhibit considerable structure: a prominent broad component with 
FWHM $\sim 3500$ $\rm km s^{-1}$, and a narrow component of $\sim 300$ $\rm 
km s^{-1}$ . Both components show a position-dependent velocity gradient along 
the conical axis with a velocity shift of $\sim 220$ $\rm km s^{-1}$ in the 
narrow component and a shift of $\sim 1500$ $\rm km s^{-1}$ in the broad component. 
Both patterns 
are consistent with radial outflow from the nuclear region. Both the continuum 
and emission lines suggest low extinction: the UV continuum is flat to the shortest 
wavelengths in the FUSE spectrum, and the best fit is a power law with virtually no 
reddening. The \ion{He}{2} 1640/1085 ratio also is reflective of a consistently low 
extinction factor across the emission-line cone. The majority of the \ion{C}{3}  and 
\ion{N}{3}  emission arises in the compact core, suggesting that the region with 
extremely high temperature is very close to the nucleus and remains unresolved. 
The line emission at $\sim 2 \arcsec$ northeast of the nucleus is strong and broad, 
characterized by (a) a strong Lya flux, but normal \ion{C}{4} flux; (b) a broad 
\ion{O}{6} line; and (c) a significantly enhanced \ion{C}{3} flux, possibly the result 
of hot knots associated with shock heating. 

\acknowledgments

W.Z. would like to thank the FUSE team, particularly W. Blair,
for their painstaking help in planning the observations of NGC 1068 and 
post-observation studies of engineering data. 

This work has been supported in part by NASA grant NAG-8-1527 and NAG-8-1133.
The FUSE and HST data presented in this paper were obtained from 
the Multimission Archive at the Space Telescope Science Institute (MAST).

\clearpage

\begin{deluxetable}{lccccl}
\tablecolumns{6}
\tablewidth{0pc}
\tablecaption{FUSE Observations of NGC 1068\label{tbl-1}}
\tablehead{
\colhead{Data Set} & \colhead{Start Time} & \colhead{Orbit} & 
\colhead{Integration Time} & \colhead{Aperture} & \colhead{Peak-up} \\
\colhead{(Pointing)} & \colhead{UT} & \colhead{} & \colhead{(sec)} & \colhead{} 
& \colhead{}
}
\startdata
P1110202000  & 2001-Nov-28 09:57:00 & 8 & 21951 & LWRS & MDRS \\
A1390201000  & 2001-Nov-29 01:23:46 & 9 & 19541 & HIRS & MDRS, HIRS\\
A1390202000  & 2001-Nov-29 16:34:46 & 9 & 19828 & HIRS & MDRS, HIRS\\
A1390203000  & 2001-Nov-30 08:38:00 & 7 & 18023 & HIRS & MDRS, HIRS\\
A1390204000  & 2001-Nov-30 20:53:34 & 9 & 17962 & HIRS & MDRS, HIRS\\
A1390205000  & 2001-Dec-01 11:52:20 & 8 & 19080 & HIRS & MDRS, HIRS\\
\enddata
\end{deluxetable}

\clearpage

\begin{deluxetable}{cccccc}
\tablecolumns{6}
\tablewidth{0pc}
\tablecaption{Flux Normalization Factor\label{tbl-2}}
\tablehead{
\colhead{Position} & \colhead{LiF1A} & \colhead{LiF2A} & 
\colhead{SiC1B} & \colhead{SiC2B} & \colhead{STIS}}
\startdata
A & 1.00 & 1.51 & 0.86 & 2.48 & 2.50 \\
B & 1.00 & 1.42 & 0.97 & 1.70 & 2.03 \\ 
C & 1.00 & 1.17 & 1.03 & 1.07 & 1.74 \\ 
D & 1.00 & 1.29 & 0.96 & 1.29 & 2.28 \\ 
E & 1.00 & 1.15 & 0.78 & 0.87 & 3.95 \\ 
F & 1.00 & 1.10 & 1.28 & 1.17 & 4.84 \\ 
G & 1.00 & 1.51 & 2.75 & 1.83 & 6.05 \\ 
\enddata
\end{deluxetable}

\clearpage
{
\begin{deluxetable}{llccc}
\tablecaption{UV Emission Lines in Spectrum with Large LWRS Aperture\label{tbl-3}}
\tablewidth{0pt}
\footnotesize
\tighttable
\tablehead{
\multicolumn{2}{c}{Emission Line} &
\colhead{Flux\tablenotemark{a}}   & \colhead{FWHM}&
\colhead{Velocity\tablenotemark{b}} 
\\
&\colhead{\AA}& \colhead{($10^{-14}$ \ergscm)} & \colhead{(\kms)}
& \colhead{(\kms)} }
\startdata
\ion{C}{3} Narrow & 977.02 & $46.2 \pm 4.0$ &$ 720 \pm 52$ & $28 \pm 16$ \\
\ion{C}{3} Broad &  & $10.1 \pm 5.3$ &$ 3489 \pm 1626$ & $1047 \pm 544$ \\
\ion{N}{3} Narrow & 990.98 & $19.8 \pm 5.2$ &$ 934 \pm 177$ & $381 \pm 47$ \\
\ion{N}{3} Broad &  & $28.5 \pm 6.5$ &$ 3393 \pm 315$ & $1371 \pm 431$ \\
\lyb\ &1025.72 & $28.4 \pm 6.5$ &$ 3392 \pm 314$ & $1340 \pm 46$ \\
\ion{O}{6} Narrow r &1037.63 & $41.1 \pm 2.5$ &$ 390 \pm 31$ & $290 \pm 6$ \\
\ion{O}{6} Narrow b & & $59.7 \pm 3.7$ &$ 340 \pm 6$ & $76 \pm 4$ \\
\ion{O}{6} Broad &1034.00 & $338 \pm 8.1$ &$ 3584 \pm 184$ & $52\pm 61$ \\
\ion{He}{2} Narrow &1085.15 & $45 \pm 2$ &$ 993 \pm 44$ & $167 \pm 16$ \\
\ion{He}{2} Broad & & $14.4 \pm 3.2$ &$ 4286 \pm 235$ & $-3912\pm 354$ \\
\enddata
\tablenotetext{a}{With $E_{B-V}=0.00$.} 
\tablenotetext{b}{With respect to the systemic redshift $z=0.0038$~\kms.}
\end{deluxetable}
}
\clearpage

{
\begin{deluxetable}{llcccl}
\tablecaption{UV Emission Lines at Position A-G \label{tbl-4}}
\tablewidth{0pt}
\footnotesize
\tighttable
\tablehead{
\multicolumn{2}{c}{Line} &
\colhead{Flux\tablenotemark{a}}   & \colhead{FWHM\tablenotemark{b}}&
\colhead{Velocity\tablenotemark{b,c}} & \colhead{Comment}
\\
&\colhead{\AA}& \colhead{($10^{-14}$ \ergscm)} & \colhead{(\kms)}
& \colhead{(\kms)} &
}
\startdata
\multicolumn{5}{l}{Position A}\\
\ion{C}{3} &  977.02 & $2.3 \pm 1.5$ &$ 564 \pm 243$ & $604 \pm 198$ \\
\ion{N}{3} Narrow &  990.98 & $2.0 \pm 1.0$ &$ 230 \pm 108$ & $140 \pm 60$ \\
\ion{N}{3} Broad &        & $0.0 \pm 0.0$ &$ 25434 \pm 92630$ & $1284 \pm 30138$ \\
\lyb\ & 1025.72 & $0.7 \pm 0.4$ &$ 878 \pm 0$ & $257 \pm 192$ \\
\ion{O}{6} Narrow & 1037.63 & $7.9 \pm 0.6$ &$ 655 \pm 38$ & $376 \pm 6$ \\
\ion{O}{6} Broad & 1034.00 & $7.8 \pm 0.8$ &$ 3131 \pm 488$ & $1192 \pm 207$ \\
\ion{He}{2} & 1085.15 & $3.1 \pm 1.0$ &$ 1098 \pm 968$ & $16 \pm 116$ \\
\lya\ Narrow & 1215.67 & $37.1 \pm 2.8$ &$ 1580 \pm 96$ & $393 \pm 37$ \\
\lya\ Broad &        & $19.2 \pm 2.1$ &$ 5283 \pm 743$ & $393 \pm 0$ \\
\ion{C}{4} Narrow & 1549.50 & $18.8 \pm 6.5$ &$ 1432 \pm 95$ & $374 \pm 96$ \\
\ion{C}{4} Broad &        & $7.6 \pm 1.4$ &$ 6178 \pm 1623$ & $507 \pm 424$ \\
\ion{He}{2} & 1640.46 & $9.8 \pm 0.7$ &$ 1858 \pm 134$ & $10 \pm 67$ \\
\ion{N}{3} & 1750.00 & $0.3 \pm 0.5$ &$ 1818 \pm 1916$ & $3691 \pm 1758$ \\
\ion{C}{3}] Narrow & 1908.73 & $6.8 \pm 1.3$ &$ 2111 \pm 228$ & $43 \pm 72$ \\
\ion{C}{3}] Broad &        & $9.4 \pm 1.3$ &$ 6883 \pm 1106$ & $-535 \pm 348$ \\
\multicolumn{5}{l}{Position B}\\
\ion{C}{3} &  977.02 & $14.1 \pm 3.5$ &$ 596 \pm 290$ & $-31 \pm 55$ \\
\ion{N}{3} Narrow &  990.98 & $8.8 \pm 3.5$ &$ 1048 \pm 351$ & $519 \pm 190$ \\
\ion{N}{3} Broad &        & $2.4 \pm 2.7$ &$ 2005 \pm 2083$ & $2332 \pm 1454$ \\
\lyb\ & 1025.72 & $5.2 \pm 1.0$ &$ 878 \pm 0$ & $-2 \pm 105$ \\
\ion{O}{6} Narrow & 1037.63 & $23.9 \pm 5.1$ &$ 385 \pm 34$ & $333 \pm 12$ \\
\ion{O}{6} Broad & 1034.00 & $123.5 \pm 2.5$ &$ 2535 \pm 35$ & $298 \pm 49$ \\
\ion{He}{2} & 1085.15 & $10.2 \pm 1.5$ &$ 1108 \pm 177$ & $-18 \pm 74$ \\
\lya\ Narrow & 1215.67 & $158.2 \pm 0.4$ &$ 1528 \pm 18$ & $366 \pm 10$ \\
\lya\ Broad &        & $119.9 \pm 7.8$ &$ 7676 \pm 787$ & $366 \pm 0$ \\
\ion{C}{4} Narrow & 1549.50 & $95.8 \pm 2.9$ &$ 1348 \pm 32$ & $284 \pm 12$ \\
\ion{C}{4} Broad &        & $60.3 \pm 3.1$ &$ 4424 \pm 172$ & $-92 \pm 66$ \\
\ion{He}{2} & 1640.46 & $47.0 \pm 1.6$ &$ 1872 \pm 66$ & $-234 \pm 31$ \\
\ion{N}{3} & 1750.00 & $4.1 \pm 1.0$ &$ 1848 \pm 569$ & $299 \pm 169$ \\
\ion{C}{3}] Narrow & 1908.73 & $29.6 \pm 2.4$ &$ 2229 \pm 95$ & $-126 \pm 27$ \\
\ion{C}{3}] Broad &        & $36.7 \pm 2.2$ &$ 6296 \pm 408$ & $-678 \pm 110$ \\
\multicolumn{5}{l}{Position C}\\
\ion{C}{3} &  977.02 & $38.6 \pm 4.3$ &$ 911 \pm 99$ & $-7 \pm 49$ \\
\ion{N}{3} Narrow &  990.98 & $11.0 \pm 10.3$ &$ 989 \pm 558$ & $-8 \pm 202$ \\
\ion{N}{3} Broad &        & $29.0 \pm 14.8$ &$ 3207 \pm 767$ & $870 \pm 770$ \\
\lyb\ & 1025.72 & $12.6 \pm 2.9$ &$ 878 \pm 132$ & $-84 \pm 61$ \\
\ion{O}{6} Narrow & 1037.63 & $46.2 \pm 6.3$ &$ 333 \pm 22$ & $290 \pm 9$ \\
\ion{O}{6} Broad & 1034.00 & $325.3 \pm 43.4$ &$ 2523 \pm 118$ & $162 \pm 55$ \\
\ion{He}{2} & 1085.15 & $17.0 \pm 1.7$ &$ 1149 \pm 116$ & $-70 \pm 55$ \\
\lya\ Narrow & 1215.67 & $278.5 \pm 4.7$ &$ 1352 \pm 21$ & $395 \pm 7$ \\
\lya\ Broad &        & $253.3 \pm 13.5$ &$ 9104 \pm 641$ & $395 \pm 0$ \\
\ion{C}{4} Narrow & 1549.50 & $167.6 \pm 9.9$ &$ 1472 \pm 59$ & $172 \pm 19$ \\
\ion{C}{4} Broad &        & $131.0 \pm 9.8$ &$ 4193 \pm 244$ & $-390 \pm 73$ \\
\ion{He}{2} & 1640.46 & $92.0 \pm 3.8$ &$ 1830 \pm 82$ & $-406 \pm 35$ \\
\ion{N}{3} & 1750.00 & $10.1 \pm 1.6$ &$ 2605 \pm 456$ & $118 \pm 160$ \\
\ion{C}{3}] Narrow & 1908.73 & $38.4 \pm 4.4$ &$ 1902 \pm 107$ & $-270 \pm 28$ \\
\ion{C}{3}] Broad &        & $80.3 \pm 4.0$ &$ 5053 \pm 219$ & $-672 \pm 56$ \\
\multicolumn{5}{l}{Position D}\\
\ion{C}{3} &  977.02 & $31.2 \pm 3.2$ &$ 900 \pm 0$ & $183 \pm 52$ \\
\ion{N}{3} Narrow &  990.98 & $11.1 \pm 3.3$ &$ 888 \pm 243$ & $372 \pm 93$ \\
\ion{N}{3} Broad &        & $44.1 \pm 71.1$ &$ 5775 \pm 3455$ & $-3804 \pm 4077$ \\
\lyb\ & 1025.72 & $6.2 \pm 1.8$ &$ 878 \pm 0$ & $-46 \pm 0$ \\
\ion{O}{6} Narrow & 1037.63 & $14.9 \pm 0.9$ &$ 380 \pm 31$ & $287 \pm 9$ \\
\ion{O}{6} Broad & 1034.00 & $181.8 \pm 17.1$ &$ 2506 \pm 126$ & $-283 \pm 32$ \\
\ion{He}{2} & 1085.15 & $15.6 \pm 1.2$ &$ 940 \pm 85$ & $21 \pm 36$ \\
\lya\ Narrow & 1215.67 & $291.4 \pm 14.3$ &$ 1248 \pm 42$ & $388 \pm 7$ \\
\lya\ Broad &        & $145.9 \pm 10.6$ &$ 5718 \pm 1708$ & $388 \pm 0$ \\
\ion{C}{4} Narrow & 1549.50 & $143.7 \pm 18.7$ &$ 1777 \pm 125$ & $6 \pm 21$ \\
\ion{C}{4} Broad &        & $105.9 \pm 18.4$ &$ 3388 \pm 130$ & $-793 \pm 17$ \\
\ion{He}{2} & 1640.46 & $88.6 \pm 3.4$ &$ 1928 \pm 73$ & $-504 \pm 35$ \\
\ion{N}{3} & 1750.00 & $9.1 \pm 2.3$ &$ 2415 \pm 404$ & $15 \pm 188$ \\
\ion{C}{3}] Narrow & 1908.73 & $45.3 \pm 2.9$ &$ 2181 \pm 493$ & $-367 \pm 114$ \\
\ion{C}{3}] Broad &        & $58.8 \pm 0.0$ &$ 5201 \pm 0$ & $-877 \pm 39538$ \\
\multicolumn{5}{l}{Position E}\\
\ion{C}{3} &  977.02 & $17.7 \pm 4.2$ &$ 1149 \pm 412$ & $27 \pm 119$ \\
\ion{N}{3} Narrow &  990.98 & $6.7 \pm 2.6$ &$ 577 \pm 177$ & $375 \pm 84$ \\
\ion{N}{3} Broad &        & $12.9 \pm 15.0$ &$ 2584 \pm 2236$ & $-764 \pm 1615$ \\
\lyb\ & 1025.72 & $3.6 \pm 1.2$ &$ 878 \pm 0$ & $68 \pm 0$ \\
\ion{O}{6} Narrow & 1037.63 & $13.0 \pm 1.2$ &$ 639 \pm 42$ & $252 \pm 17$ \\
\ion{O}{6} Broad & 1034.00 & $52.9 \pm 2.9$ &$ 3145 \pm 168$ & $-475 \pm 78$ \\
\ion{He}{2} & 1085.15 & $12.1 \pm 0.8$ &$ 818 \pm 58$ & $208 \pm 25$ \\
\lya\ Narrow & 1215.67 & $359.0 \pm 9.4$ &$ 1265 \pm 24$ & $474 \pm 7$ \\
\lya\ Broad &        & $72.4 \pm 6.4$ &$ 4953 \pm 1166$ & $474 \pm 0$ \\
\ion{C}{4} Narrow & 1549.50 & $107.2 \pm 10.9$ &$ 1812 \pm 141$ & $79 \pm 71$ \\
\ion{C}{4} Broad &        & $63.5 \pm 11.0$ &$ 3241 \pm 134$ & $-981 \pm 99$ \\
\ion{He}{2} & 1640.46 & $64.0 \pm 2.9$ &$ 4490 \pm 157$ & $1112 \pm 5$ \\
\ion{N}{3} & 1750.00 & $9.1 \pm 1.6$ &$ 2986 \pm 549$ & $362 \pm 237$ \\
\ion{C}{3}] Narrow & 1908.73 & $30.3 \pm 4.2$ &$ 2185 \pm 135$ & $82 \pm 55$ \\
\ion{C}{3}] Broad &        & $39.3 \pm 4.3$ &$ 4489 \pm 251$ & $-1026 \pm 162$ \\
\multicolumn{5}{l}{Position F}\\
\ion{C}{3} &  977.02 & $15.8 \pm 4.3$ &$ 900 \pm 270$ & $424 \pm 119$ \\
\ion{N}{3} Narrow &  990.98 & $7.1 \pm 5.8$ &$ 1056 \pm 625$ & $164 \pm 289$ \\
\ion{N}{3} Broad &        & $8.0 \pm 9.4$ &$ 3540 \pm 1540$ & $333 \pm 1681$ \\
\lyb\ & 1025.72 & $9.0 \pm 0.7$ &$ 878 \pm 0$ & $257 \pm 47$ \\
\ion{O}{6} Narrow & 1037.63 & $11.4 \pm 1.9$ &$ 219 \pm 30$ & $152 \pm 17$ \\
\ion{O}{6} Broad & 1034.00 & $96.1 \pm 15.6$ &$ 1785 \pm 16$ & $456 \pm 20$ \\
\ion{He}{2} & 1085.15 & $8.3 \pm 0.7$ &$ 665 \pm 66$ & $266 \pm 25$ \\
\lya\ Narrow & 1215.67 & $299.6 \pm 9.6$ &$ 1223 \pm 33$ & $520 \pm 10$ \\
\lya\ Broad &        & $46.8 \pm 7.4$ &$ 4582 \pm 1014$ & $520 \pm 0$ \\
\ion{C}{4} Narrow & 1549.50 & $73.3 \pm 5.9$ &$ 1181 \pm 72$ & $313 \pm 29$ \\
\ion{C}{4} Broad &        & $34.7 \pm 5.3$ &$ 3365 \pm 323$ & $-461 \pm 178$ \\
\ion{He}{2} & 1640.46 & $42.6 \pm 2.8$ &$ 1132 \pm 74$ & $305 \pm 33$ \\
\ion{N}{3} & 1750.00 & $4.7 \pm 1.6$ &$ 1289 \pm 702$ & $1040 \pm 211$ \\
\ion{C}{3}] Narrow & 1908.73 & $29.1 \pm 1.2$ &$ 1500 \pm 77$ & $400 \pm 41$ \\
\ion{C}{3}] Broad &        & $14.3 \pm 0.9$ &$ 3848 \pm 392$ & $-1679 \pm 139$ \\
\multicolumn{5}{l}{Position G}\\
\ion{C}{3} &  977.02 & $12.5 \pm 5.4$ &$ 344 \pm 133$ & $91 \pm 80$ \\
\ion{N}{3} Narrow &  990.98 & $8.7 \pm 4.2$ &$ 833 \pm 398$ & $417 \pm 148$ \\
\ion{N}{3} Broad &        & $10.8 \pm 58.4$ &$ 3237 \pm 7653$ & $-2085 \pm 9897$ \\
\lyb\ & 1025.72 & $4.3 \pm 0.9$ &$ 878 \pm 0$ & $-61 \pm 93$ \\
\ion{O}{6} Narrow & 1037.63 & $13.3 \pm 2.6$ &$ 372 \pm 16$ & $189 \pm 14$ \\
\ion{O}{6} Broad & 1034.00 & $37.6 \pm 4.9$ &$ 2852 \pm 422$ & $416 \pm 95$ \\
\ion{He}{2} & 1085.15 & $7.5 \pm 1.4$ &$ 617 \pm 105$ & $150 \pm 58$ \\
\lya\ Narrow & 1215.67 & $152.2 \pm 5.8$ &$ 1208 \pm 35$ & $523 \pm 12$ \\
\lya\ Broad &        & $32.2 \pm 3.9$ &$ 4627 \pm 848$ & $523 \pm 0$ \\
\ion{C}{4} Narrow & 1549.50 & $50.8 \pm 3.0$ &$ 1227 \pm 62$ & $251 \pm 27$ \\
\ion{C}{4} Broad &        & $19.5 \pm 2.3$ &$ 4877 \pm 581$ & $-283 \pm 216$ \\
\ion{He}{2} & 1640.46 & $28.0 \pm 1.7$ &$ 1047 \pm 58$ & $265 \pm 29$ \\
\ion{N}{3} & 1750.00 & $1.5 \pm 0.8$ &$ 708 \pm 334$ & $1453 \pm 195$ \\
\ion{C}{3}] Narrow & 1908.73 & $12.5 \pm 3.1$ &$ 1797 \pm 252$ & $333 \pm 83$ \\
\ion{C}{3}] Broad &        & $12.4 \pm 3.5$ &$ 4491 \pm 658$ & $-870 \pm 510$ \\
\enddata
\tablenotetext{a}{With $E_{B-V}=0.00$.}
\tablenotetext{b}{Values with zero errors are pre-fixed.}
\tablenotetext{c}{With respect to the systemic redshift $z=0.0038$ \kms.}
\end{deluxetable}
}
\clearpage

\centerline{\bf Figure Captions}
\bigskip

\figcaption{HST image of the central region of NGC 1068 obtained with 
WFPC2/F218W, with a field of view of $\sim 25$\arcsec. 
The north is up, east is to the left, and the nucleus is marked with a cross. 
Seven FUSE slit positions 
are marked with letters A-G, while C is at the nucleus. Note that the actual 
slit width is slightly wider than that in the illustration. The narrow slits 
along the near-perpendicular direction are that of HST/STIS: The three 
parallel slits along the northeast and southwest direction are 
0\farcs 2 wide \citep[STIS-A,][]{cecil}, and the single long slit is another 
that is 0\farcs 1 wide \citep[STIS-B, ][]{krm}. 
\label{fig1}}

\figcaption{FUSE spectrum of NGC 1068 at full aperture (30\arcsec\ square). 
The data are binned to 0.1 \AA. Propagation errors are plotted in the lower 
panel. The effect of major geocoronal emission lines are marked with the 
Earth symbols, and data at these wavelength bins are removed.
\label{fig2}}

\figcaption{Typical relative drifts of the FUSE SiC2 channel vs. orbital time 
(after orbital sunset), with respect to the LiF1 channel. Data are collected 
from various orbits with occultation.
\label{fig3}}

\figcaption{Count rate vs. orbital time in a sample data set A1390201001. 
Different channels register their peaks at different times, which are 
presumed to coincide with the nucleus position. 
Incomplete segments are not used (Position A in SiC channels and position G 
in LiF channels). 
\label{fig4}}

\figcaption{Merged FUSE spectra at seven FUSE slit positions, binned to 0.5 
\AA. The nucleus is at position C. 
\label{fig5}}

\figcaption{STIS spectra extracted at different windows that correspond 
to the respective FUSE slits. Fluxes are normalized to the corresponding FUSE data.
\label{fig6}}

\figcaption{Line profiles at different FUSE slit positions. The lines are 
redshifted \ovi\
\lm 1037 (centered at 0 \kms), \civ\ \lm 1549, \niii\ \lm 991, and \ciii\ 
\lm 977, respectively.
\label{fig7}}

\figcaption{Intensities of major emission lines at different FUSE positions, 
from STIS (left panels) and FUSE (right panels). When a line is fitted with 
multiple components with reasonably accuracy, the flux presented is their sum.
For \ciii\ \lm 977, only one narrow component is fitted.
\label{fig8}}

\figcaption{Emission and absorption lines near the \ion{O}{6} wavelength, 
from position A to G
after subtraction of the fitted continuum. The spectrum marked ``All'' in the 
lower bottom panel is from the data with a large LWRS aperture.
The dashed lines mark the wavelengths of redshifted Ly$\beta$, \ion{O}{6} 
\lm\lm 1031.95/1037.63, respectively, from left to right.
\label{fig9}}

\figcaption{Properties of fitted \ion{O}{6} emission line at different FUSE slit 
positions. Line fluxes are in units of $10^{-14}$ \ergscm, FWHM and velocity 
are in units of \kms.
\label{fig10}}

\figcaption{Properties of fitted \ion{C}{4} emission line (from HST/STIS) at 
different FUSE positions. Line fluxes are in units of $10^{-14}$ \ergscm, 
FWHM and velocity 
are in units of \kms.
\label{fig11}}

\figcaption{Line ratios at different slit positions. 
The fitting results of \civ\ 
and \ovi\ are for their narrow-line components.
\label{fig12}}

\clearpage

\setcounter{figure}{0}
\begin{figure}
\plotone{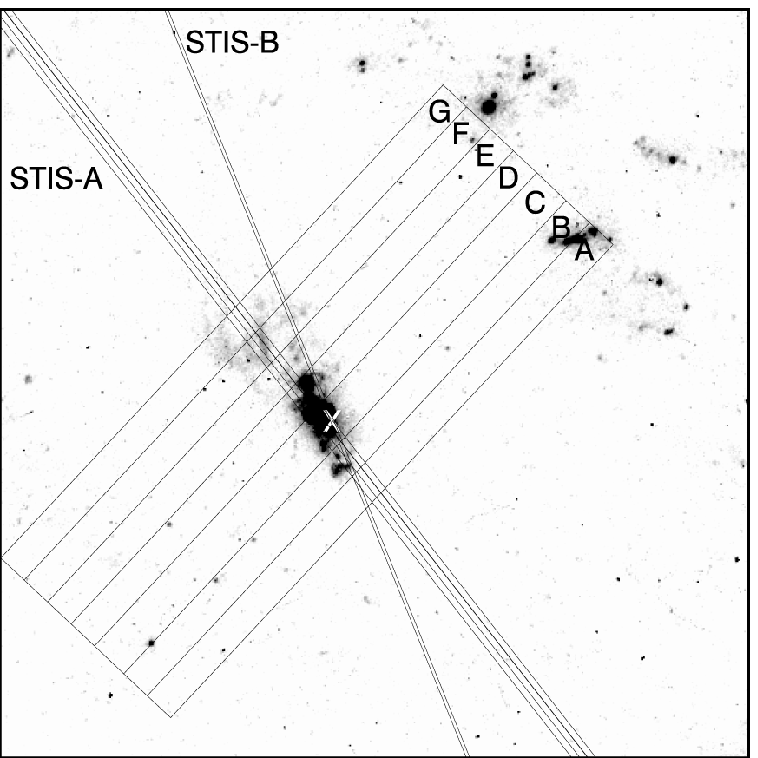}
\caption{~}
\end{figure}
\clearpage

\begin{figure}
\plotone{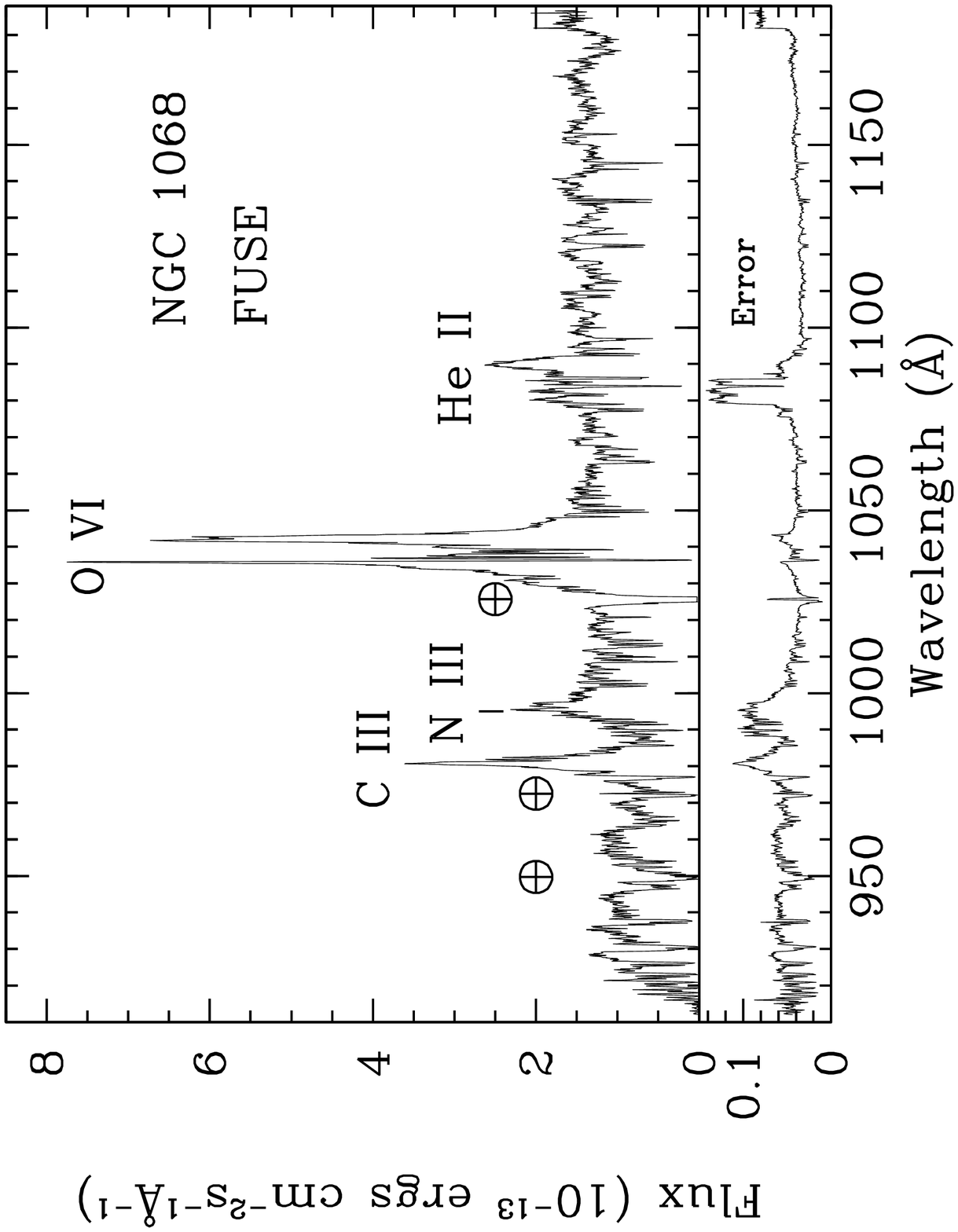}
\caption{~}
\end{figure}
\clearpage

\begin{figure}
\plotone{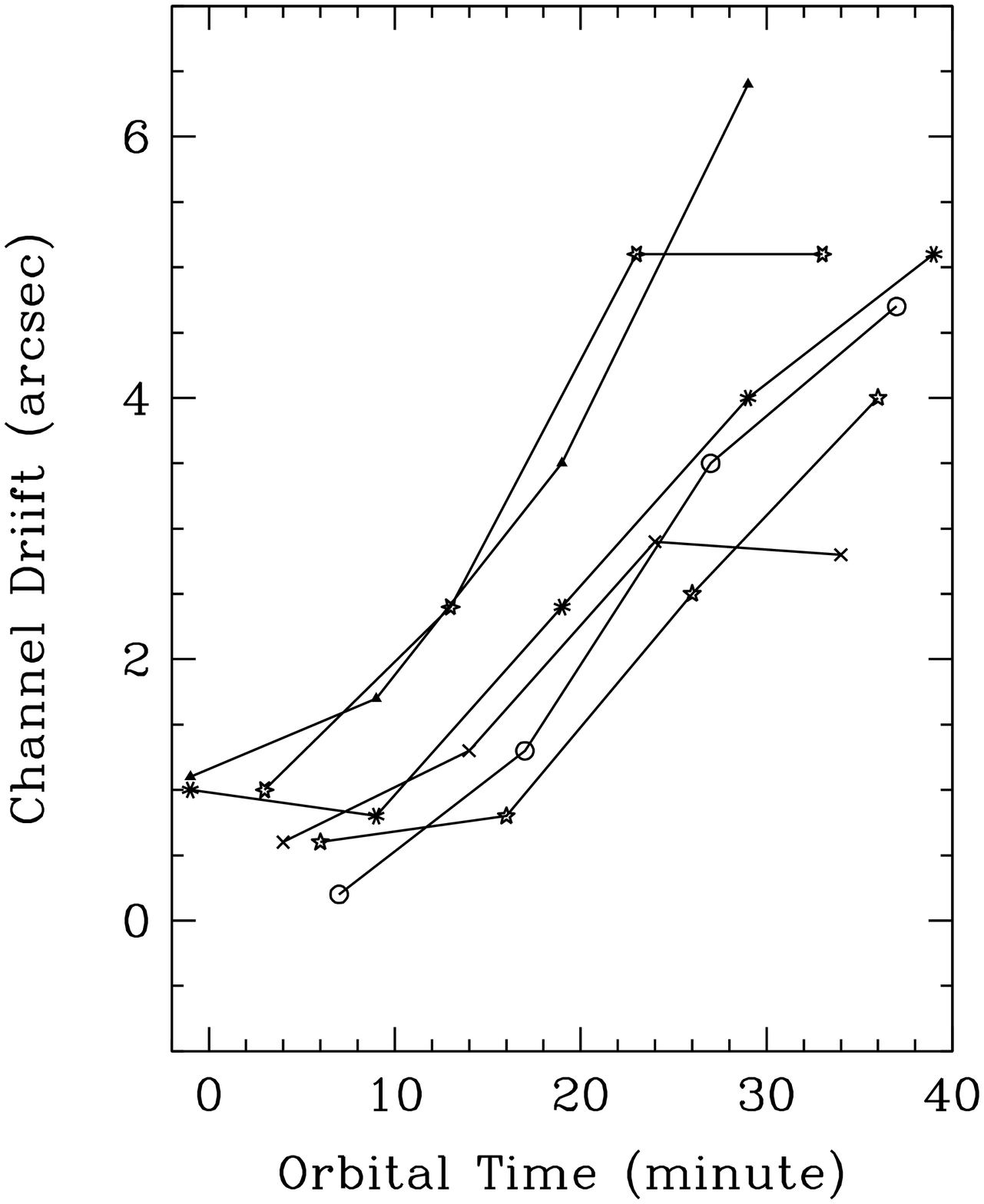}
\caption{~}
\end{figure}
\clearpage

\begin{figure}
\plotone{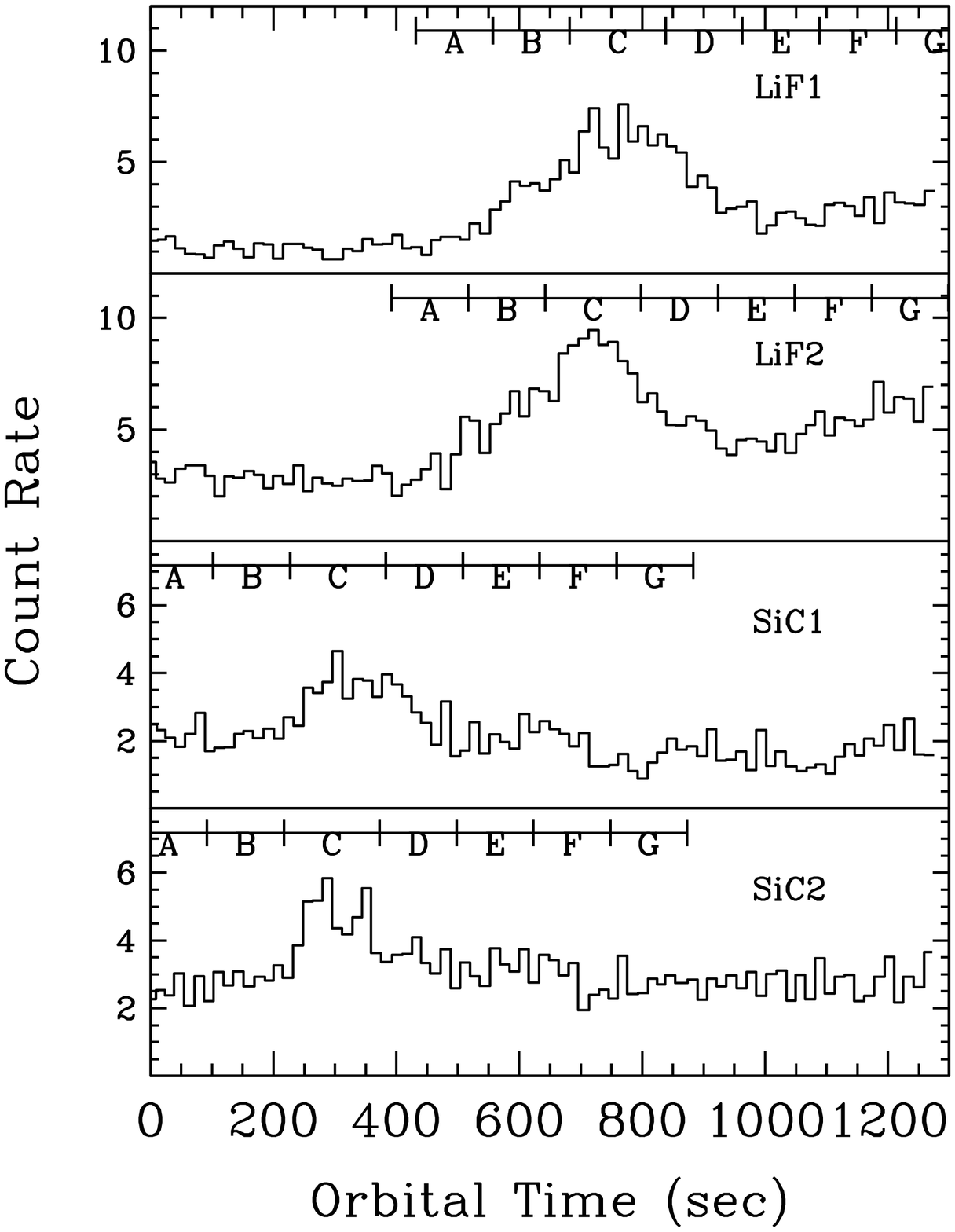}
\caption{~}
\end{figure}

\begin{figure}
\plotone{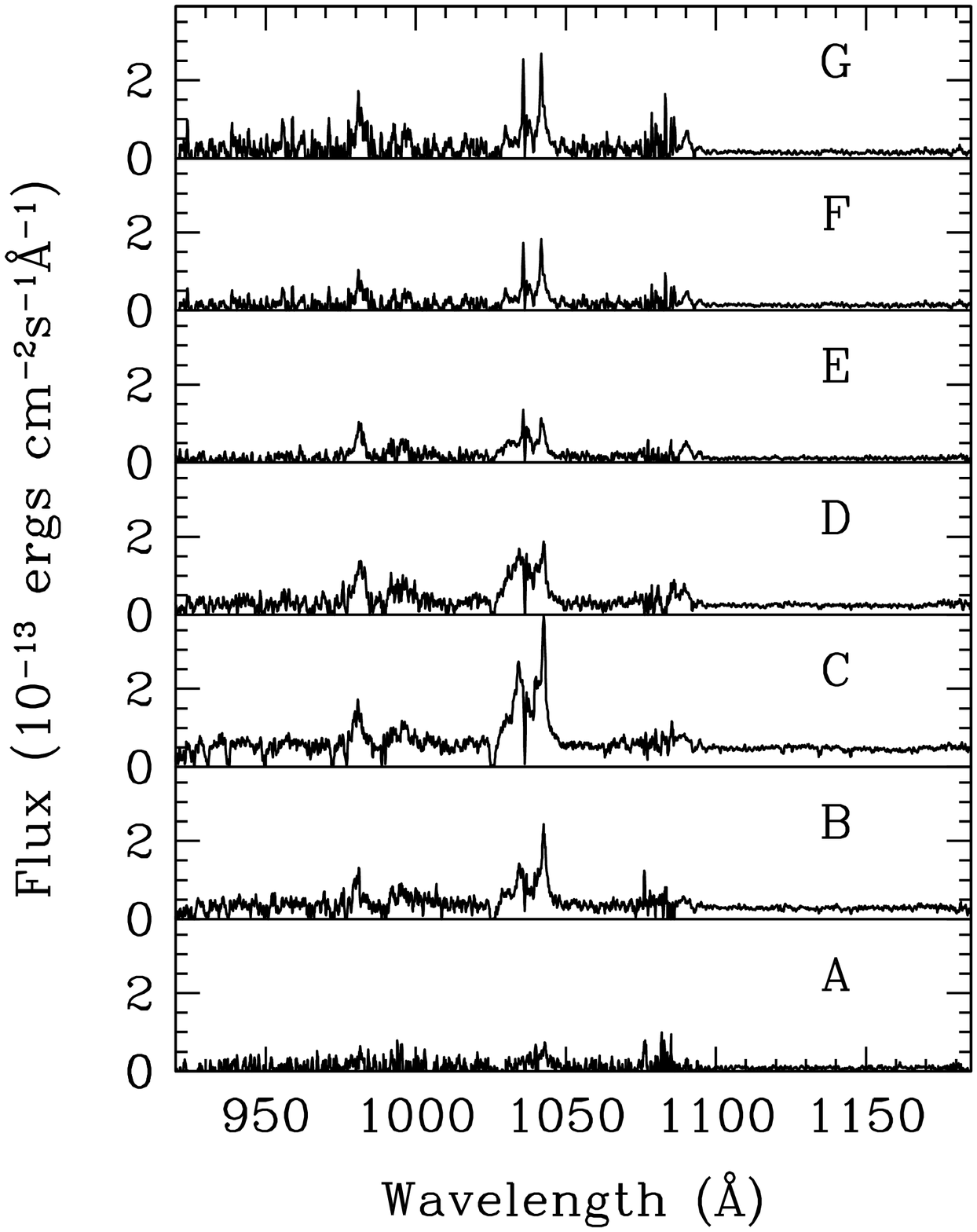}
\caption{~}
\end{figure}

\clearpage
\begin{figure}
\plotone{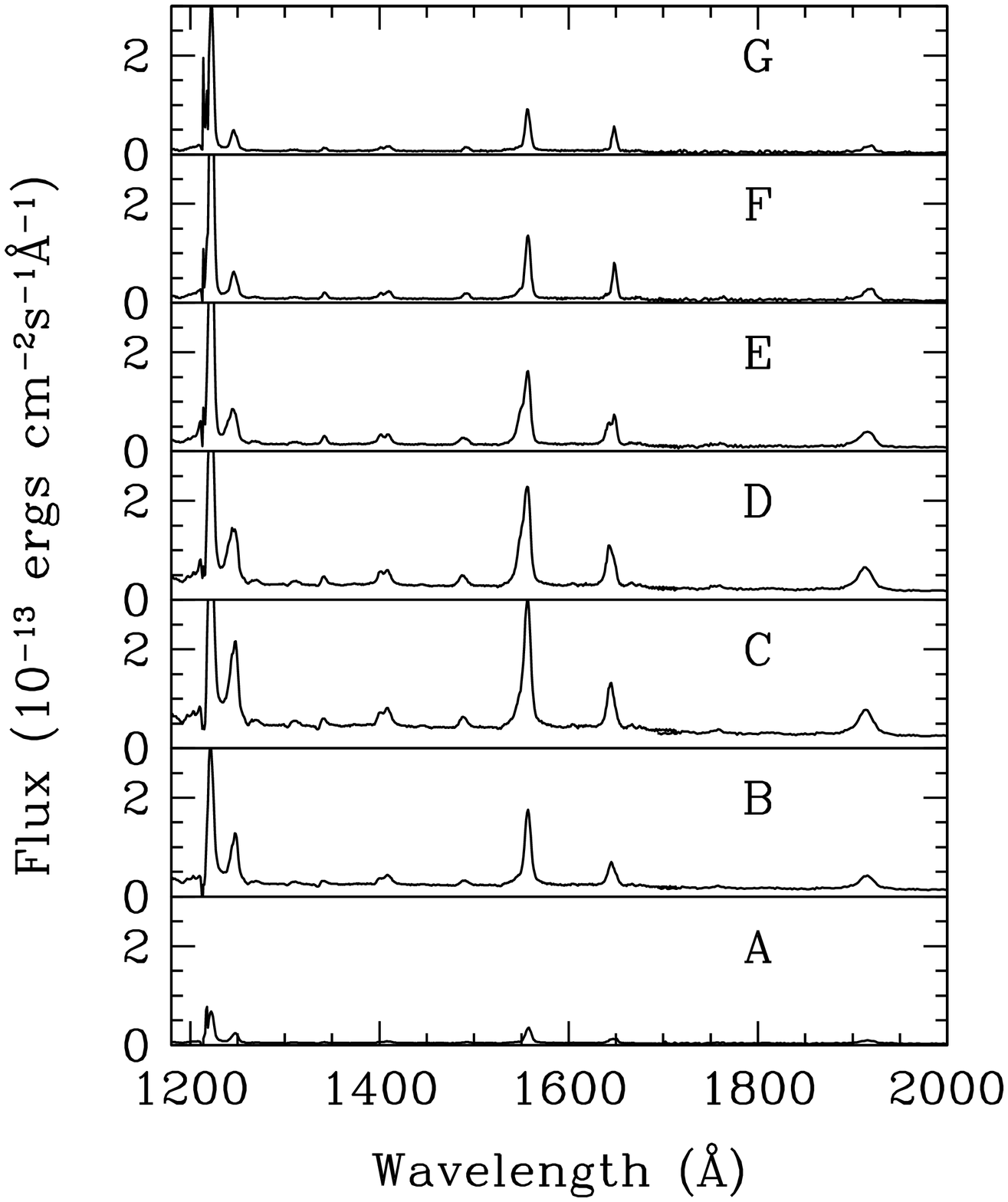}
\caption{~}
\end{figure}

\clearpage
\begin{figure}
\plotone{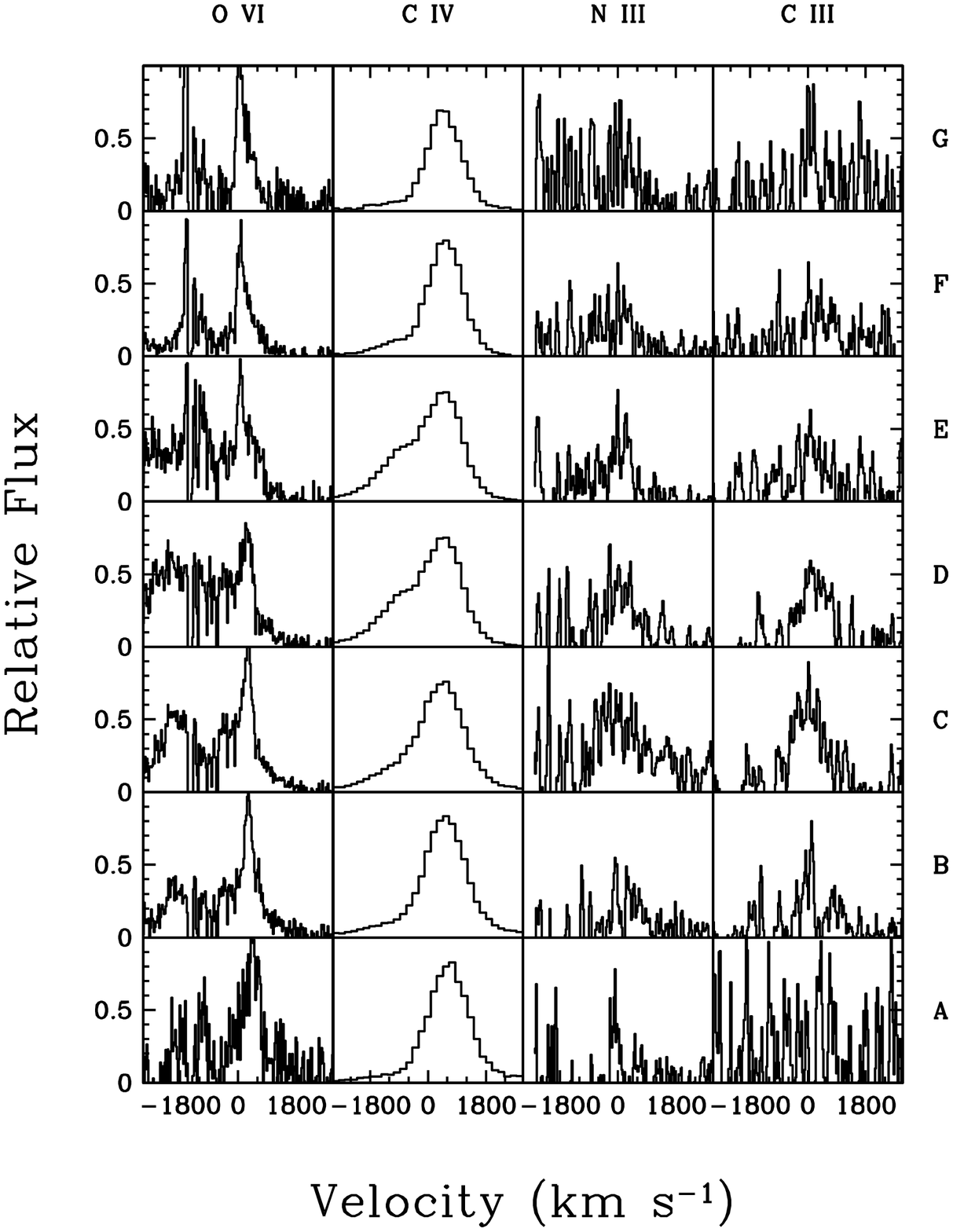}
\caption{~}
\end{figure}

\clearpage
\begin{figure}
\plotone{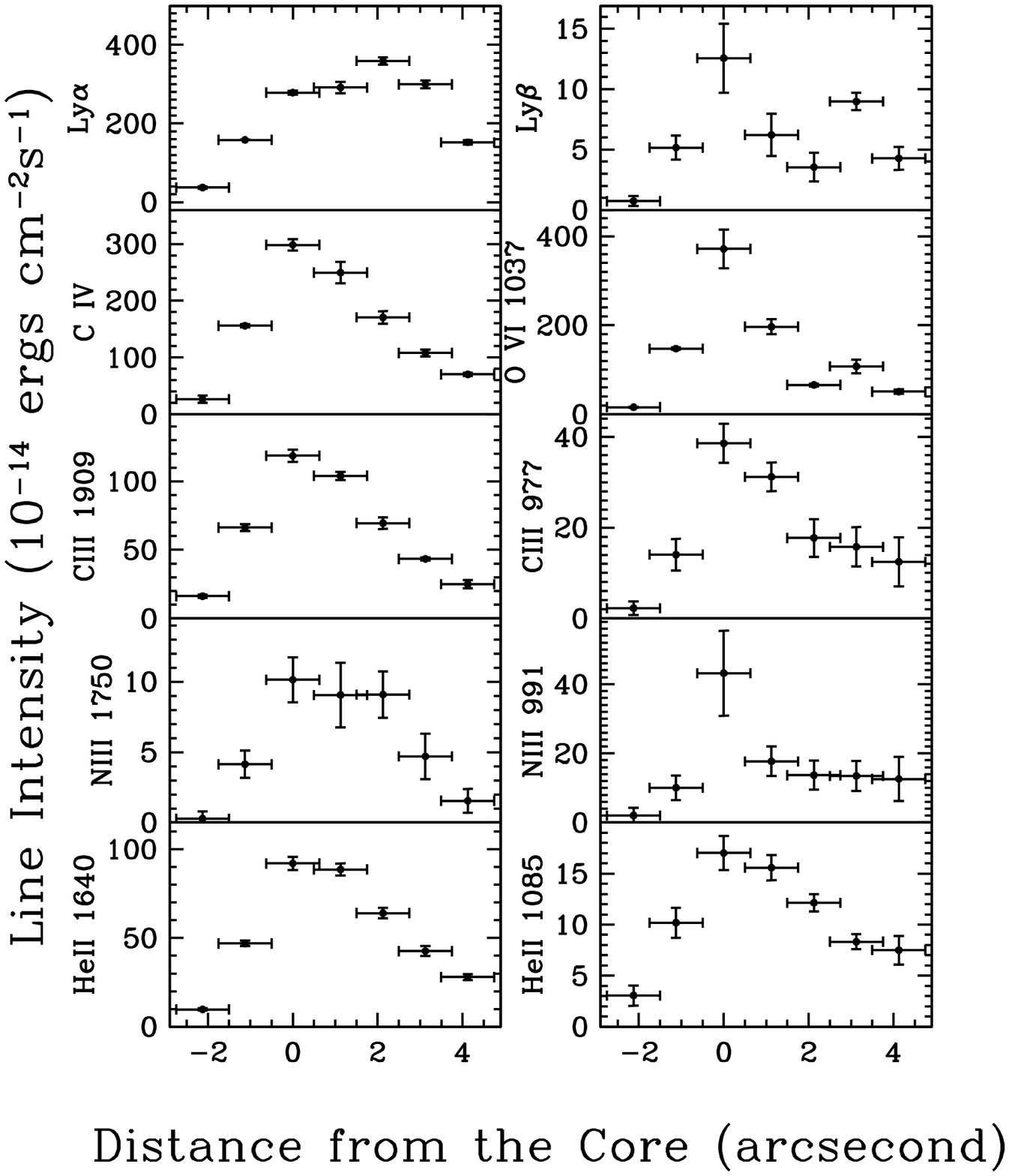}
\caption{~}
\end{figure}

\clearpage
\begin{figure}
\plotone{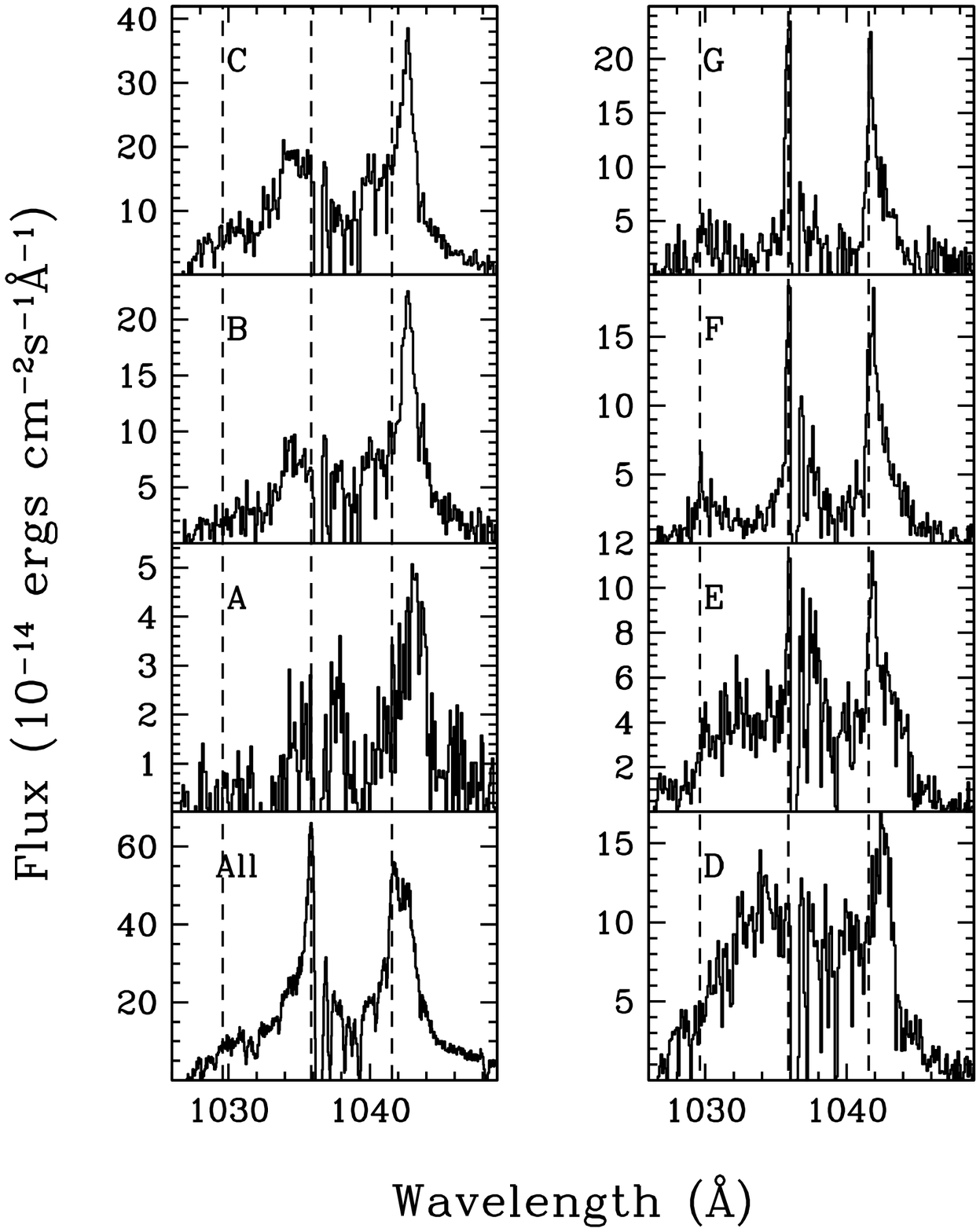}
\caption{~}
\end{figure}

\clearpage
\begin{figure}
\plotone{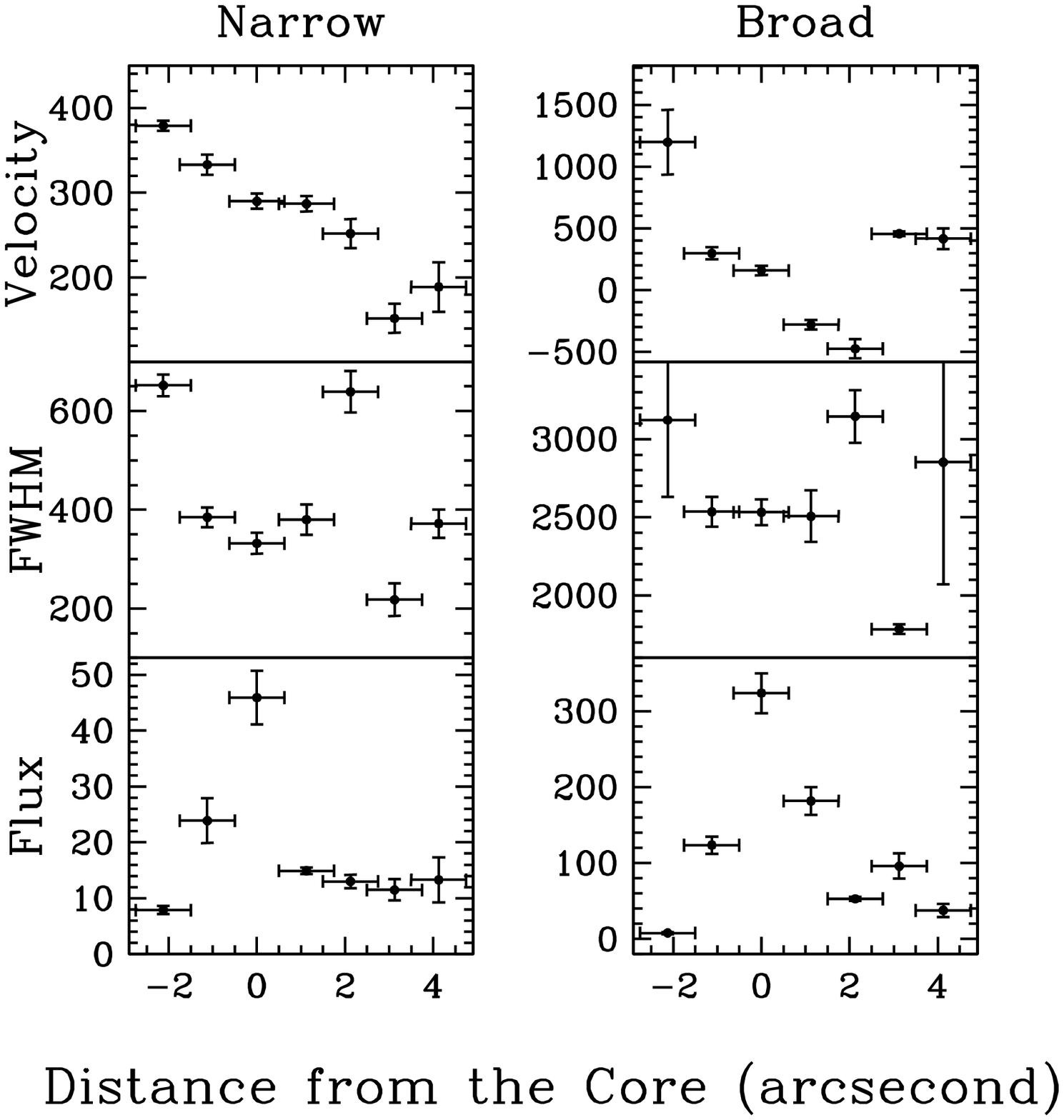}
\caption{~}
\end{figure}

\clearpage
\begin{figure}
\plotone{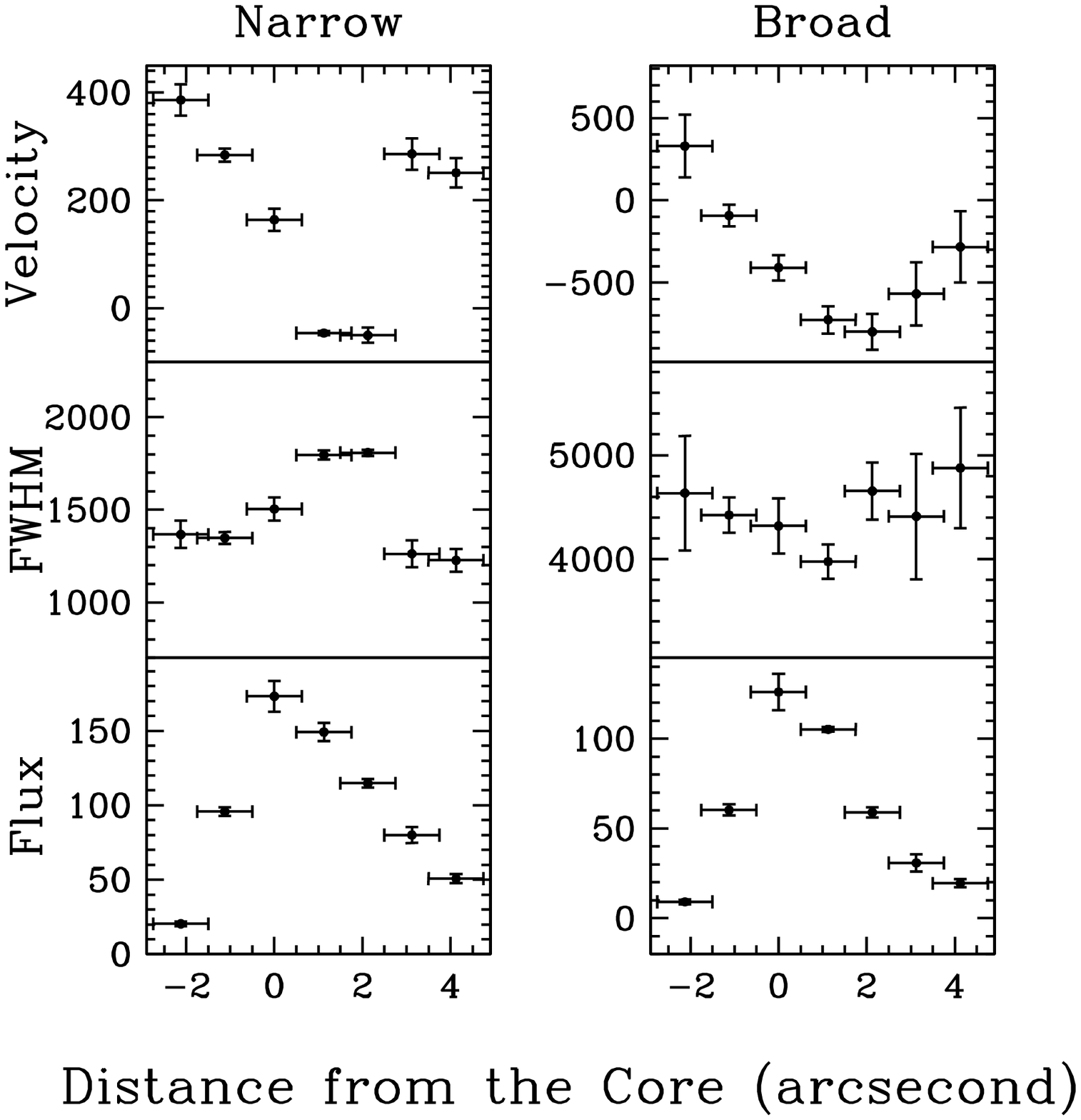}
\caption{~}
\end{figure}

\clearpage
\begin{figure}
\plotone{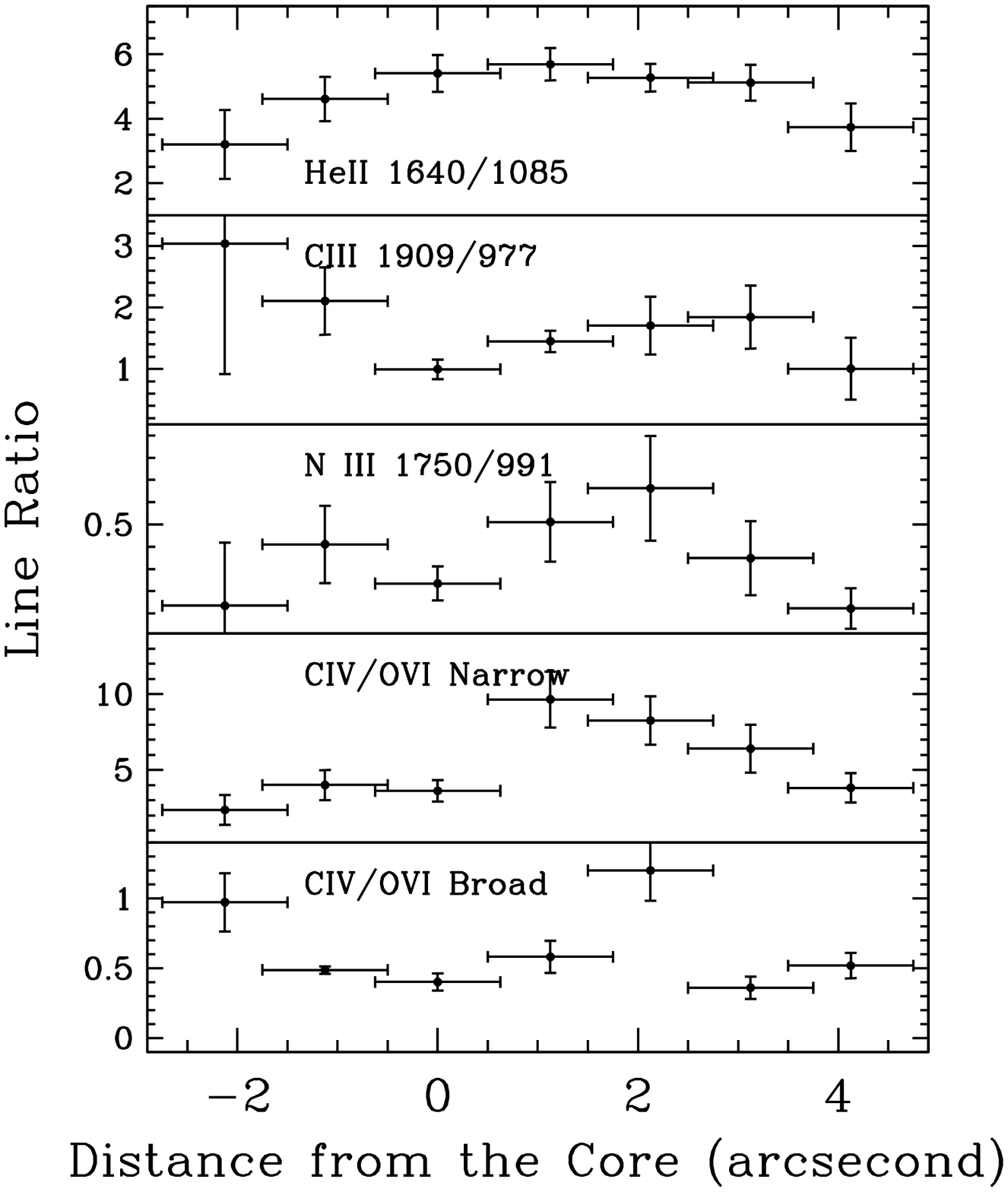}
\caption{~}
\end{figure}

\end{document}